\documentclass[5p, authoryear]{elsarticle}

\usepackage[latin1]{inputenc}
\usepackage{fontenc}
\usepackage{graphicx}

\newcommand{\ud}{\rm d}

\begin{document}

\begin{frontmatter}

\title{Hydrogen Balmer Emission Lines and the Complex Broad Line
  Region Structure}

\author[dap]{G. La Mura\corref{cor1}}
\ead{giovanni.lamura@unipd.it}
\author[dap]{F. Di Mille}
\ead{francesco.dimille@unipd.it}
\author[aob,inic]{L. \v C. Popovi\'c}
\ead{lpopovic@aob.bg.ac.yu}
\author[dap]{S. Ciroi}
\ead{stefano.ciroi@unipd.it}
\author[dap]{P. Rafanelli}
\ead{piero.rafanelli@unipd.it}
\author[mbg]{D. Ili\'c}
\ead{dilic@matf.bg.ac.yu}

\address[dap]{Department of Astronomy, University of Padova, Vicolo
  dell'Osservatorio 3, I-35122 Padova, Italy}
\address[aob]{Astronomical Observatory, Volgina 7, 11060 Belgrade,
  Serbia}
\address[inic]{Isaac Newton Institute of Chile, Yugoslavia Branch,
  11060 Belgrade, Serbia}
\address[mbg]{Department of Astronomy, Faculty of Mathematics, University of
  Belgrade, Studentski trg 16, 11000 Belgrade, Serbia}

\cortext[cor1]{Corresponding author (phone number +39 3391845700)}

\begin{abstract}
In this work we investigate the properties of the broad emission line
components in the Balmer series of a sample of Type 1 Active Galactic Nuclei
(AGN). Using the Boltzmann Plot method as a diagnostic tool for physical
conditions in the plasma, we detect a relationship among the kinematical and
thermo-dynamical properties of these objects. In order to further clarify the
influence of the central engines on the surrounding material, we look for
signatures of structure in the broad line emitting regions, that could affect
the optical domain of the observed spectra. Using a combination of line
profile analysis and kinematical modeling of the emitting plasma, we study how
the emission line broadening functions are influenced by different structural
configurations. The observed profiles are consistent with flattened structures
seen at quite low inclinations, typically with $i < 20$°. Since this result is
in good agreement with some independent observations at radio frequencies, we
apply a new formalism to study the properties of AGN central engines.
\end{abstract}

\end{frontmatter}

\section{Introduction}
The spectrum of a Type 1 AGN is characterized by the presence of
prominent broad emission lines, with a full width at half the maximum
(FWHM) which corresponds to velocity fields exceeding $10^3\,{\rm
  km\, s}^{-1}$. The properties of the emission lines are connected to
the physical conditions within the source and they suggest that the
lines are mainly originated by electron - ion recombinations in a
photo-ionized plasma \citep[see e. g.][]{Osterbrock89}. Unfortunately,
since the broad components of these spectral features are originated
very close to the center of the source, the structure of the so called
Broad Line Region (BLR) cannot be spatially resolved by our
observations. This limit affects our ability to investigate the
physics of AGN and to provide a self-consistent theoretical
interpretation of their observational properties.

At present it is widely accepted that AGN are powered by matter
accreting into the gravitational field of a {\it Super Massive Black
  Hole} (SMBH), since the gravitational binding energy of the accretion
flow may provide the ionizing radiation field, required to drive the
observed line emission from the surrounding plasma. Assuming the BLR
to be in virial equilibrium under the dynamical influence of this
central engine \citep[see, for instance,][etc.]{Peterson99, Woo02,
  Sulentic06} the mass of the black hole might be estimated with an
expression in the form of:
$$M_{\rm BH} = f \cdot \frac{R_{\rm BLR}\, \Delta v^2}{G}, \eqno(1)$$
where $G$ is the gravitational constant, $R_{\rm BLR}$ an estimate of
the BLR extension, and $\Delta v$ a measurement of the gas radial
velocity distribution. The factor $f$ accounts for the unknown
structure of the gas motion pattern. It is required in order to
estimate the actual kinematical properties from the line of sight
velocity distribution, that we measure in the spectra, and it is
therefore referred to as the {\it geometrical factor}.

Many works deal with the relationship among the emission line profiles
and the actual dynamics of the BLR \citep[see, for
  instance,][]{Capriotti80, Capriotti81, Netzer90, Ferland92}. Recent
investigations, such as those of \citet{Vestergaard00} and
\citet{Nikolajuk05}, point towards a considerably flattened geometry,
which is consistent with the existence of an accretion disk around the
SMBH. In this progress report, we describe the analysis which we
performed on a sample of Type 1 AGN spectra, in order to examine the
influence of the AGN central engines onto the surrounding BLR plasma. With a
combination of observations, analytical calculations and theoretical models,
we look for indications of the BLR structural properties in the emission line
profiles. We organize this report as follows: in \S2 we introduce the relation
among BLR kinematical and thermo-dynamical properties; in \S3 we outline the
analytical formalism, that we exploit to investigate the BLR structure; in
\S4 we discuss the predictions of our BLR structural models; finally \S5 is
devoted to the discussion of the results achieved by comparing models
and observations, with a mention to some possible tests and to the
implications related to the use of spectroscopic measurements in our
sample of AGN.

\begin{figure}[t]
\begin{center}
\includegraphics[width = 8.6cm]{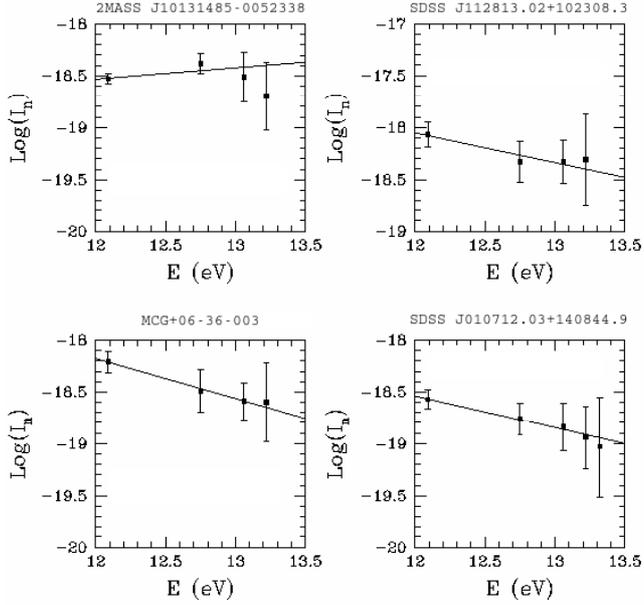}
\end{center}
\caption{Four examples of Boltzmann Plots applied to the spectra of the BLR in
  some Type 1 AGN: in the upper left panel no straight line fit is achieved
  (class iv); in the upper right panel only a poor fit, with a
  relative uncertainty $\Delta A / A > 0.2$ in the temperature
  parameter, is performed (class iii); the bottom left panel gives a
  good fit, but H$\epsilon$ could not be detected in the spectrum
  (class ii); finally the bottom right panel shows a straight line fit
  to the normalized intensities of the Balmer series up to H$\epsilon$
  (class i).}
\end{figure}
\section{Plasma physics with the Boltzmann Plots}
Due to its high density and strong interactions with the ionizing radiation
field, the BLR plasma cannot be studied by means of the standard spectroscopic
techniques, that are commonly adopted in other low density nebular
environments. This characteristic forces us to adopt alternative techniques to
investigate its properties. Looking at the flux in the broad component of an
emission line, it is possible to introduce a {\it normalized line intensity},
with respect to the atomic constants which characterize the corresponding
transition, in the form of:
$$I_{\rm n} = \frac{\lambda_{ul}\, F_{ul}}{g_u\, A_{ul}}, \eqno(2)$$
where $\lambda_{ul}$ is the line wavelength for a transition from an upper
level $u$ to a lower level $l$, $F_{ul}$ is the observed flux, $A_{ul}$ the
spontaneous transition probability, and $g_u$ the upper level's statistical
weight. Looking at the normalized emission line intensities of some spectral
features, belonging to the same transition series, it can be shown that if the
plasma approaches a condition of Local Thermo-dynamical Equilibrium (LTE), we
get \citep{Popovic03, Popovic06}:
$$\log I_{\rm n} = B - A\, E_u, \eqno(3)$$
where $E_u$ represents the upper level's excitation energy for the transition,
$B$ is a constant for all the emission lines in the series, while $A$ takes
the role of a {\it temperature parameter}, corresponding to:
$$A = \log e / (k_{\rm B} T_e), \eqno(4)$$
with $k_{\rm B}$ representing the Boltzmann constant and $T_e$ an
average value of the electron temperature in the plasma.

\begin{figure}[t]
\begin{center}
\includegraphics[width = 8.6cm]{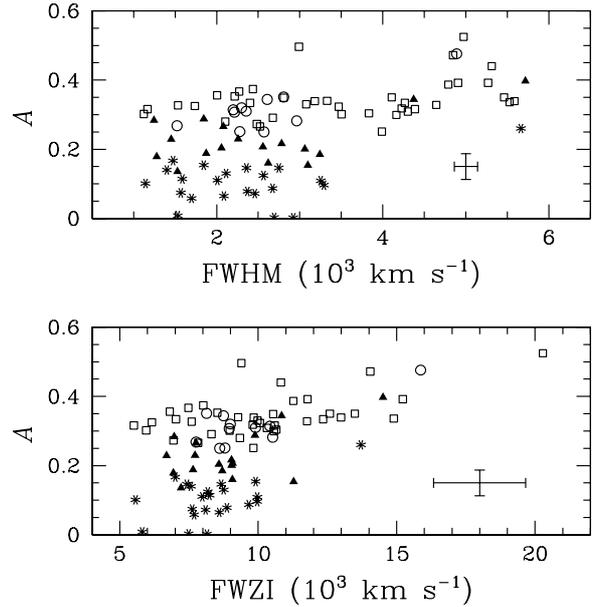}
\end{center}
\caption{$A$ parameter as a function of the BLR velocity fields, inferred by
  both the FWHM and FWZI of the Balmer lines: the open circles are sources of
  class i; the open squares represent objects of class ii; filled triangles
  are spectra of class iii and, finally, asterisks are objects with BP class
  iv. The crosses in the lower right corners of each panel are the median
  uncertainties of measurements.}
\end{figure}
The diagrams which show the emission line normalized intensities of the Balmer
series as a function of the upper level's excitation energies are known as
Boltzmann Plots (BP). A strict interpretation of the BP results, however,
requires that the physical conditions in the line emitting region should not
change abruptly. Otherwise the simple formulation of the emission line
intensities in the series would not be possible and the LTE approximation
would subsequently fail. In Fig.~1 we give some examples, showing four
different situations, that we commonly observe, applying the BP to the BLR
plasma of some AGN \citep{LaMura07}. Extracting the BLR signal from
our spectra, we computed the emission line normalized intensities of
the Balmer series as in Eq.~(2). We performed several flux
measurements, that were subsequently averaged together and compared
with the spectral noise fluctuations, in order to estimate the mean
values and their uncertainties, therefore assuming the error of the BP
to be given by: 
$$\Delta(\log I_{\rm n}) = \frac{3\, \sigma_{Ful}}{F_{ul}}, \eqno(5)$$
with $\sigma_{Ful}$ representing the emission line flux dispersion. Repeating
the emission line flux estimates was a fundamental step, because of the large
difficulties associated with flux measurements in the broad line components,
in a spectral region where several contributions to the observed emission are
blended together.

The temperature parameters extracted from our BP and illustrated in
Fig.~2 appear to be related to the emission line profiles, in the sense that
higher values of $A$ and better realizations of the prediction of Eq.~(3), are
commonly found in the range of broad line emitting objects. Unfortunately our
sample is not particularly representative in this region, but the circumstance
suggests that the thermo-dynamical properties of sources with various emission
line widths are under the influence of different ionizing radiation fields, as
it would be expected when powering the central engines of these AGN with
different combinations of SMBH mass and accretion rate. In order to improve
our understanding of the role played by these fundamental parameters, we
further investigated the processes which lead to the formation of the observed
line profiles and their relations with the actual source's dynamical
properties.

\section{Balmer emission line profile analysis}
%

\subsection{Emission line broadening from cross-correlation}
The BLR spectrum shows several emission lines corresponding to many
permitted and some semi-forbidden transitions of variously ionized
atomic species. In the optical domain, the Balmer series of hydrogen
provides a set of bright emission lines belonging to a well defined
family of interactions among matter and radiation, whose profiles are
influenced by the kinematics of gas.

To perform our study, we adopted the cross-correlation formalism,
originally described by \citet{Tonry79} and then updated by
\citet{Statler95}. As a starting point, we can approximate the spectrum of a
broad line emitting source as the convolution of an appropriate template of
narrow emission lines $T(x)$ with the corresponding broadening function
$B(x)$:
$$S(x) \simeq T(x) \ast B(x), \eqno(6)$$
where $S(x)$ is the observed spectrum, while $x$ represents a
logarithmic wavelength coordinate of the form $x = A \ln\lambda + B$,
such that the effect of radial velocities results in linear shifts
along $x$. Introducing the cross-correlation function of the spectrum
with the template:
$$X(x) = S(x) \otimes T(x) = \int S(x) T(x + x')\, \ud x', \eqno(7)$$
it can be shown that the cross-correlation function (XCF) approximates
the convolution among the template's autocorrelation function (ACF)
and the kinematical broadening function (BF) of the object
\citep{Statler95, LaMura09}:
$$X(x) \simeq [T(x) \otimes T(x)] \ast B(x). \eqno(8)$$
Since $T(x)$ is known and $X(x)$ is drawn from observations, as far as
the template is correct, it is possible to recover $B(x)$. 

Restricting our analysis to the primary cross-correlation peak, which
carries most of the kinematical information and it is weakly affected
by template mismatch, Eq.~(8) has a discrete form that, using the
simplified notation $F_i = F(x_i)$, becomes: 
$$X_k \simeq \sum_{i = 0}^N \left(\sum_{j = 0}^N T_i\, T_{i + j}\right)
B_{k - i}. \eqno (9)$$
Provided that all the functions are null when they are computed
outside the range $0 \leq i \leq N$, Eq.~(9) defines a system of
$N + 1$ linear equations in the $N + 1$ variables $B_{k - i}$ ($k \geq
i$). A standard $\chi^2$-minimization routine can be therefore used to
solve the problem.

\subsection{Analytical expressions for the broadening functions}
The BLR emission lines are influenced by the effects of complex
kinematics within the source and of radiation transfer from the source
to the observer. For this reason it is hardly conceivable that a
simple analytic expression might be used to fit the resulting
profiles. In the simple case of a random motion distribution, the
emission line profiles would approximately match a Gaussian function,
but, in presence of ordered kinematical components, we expect
significant deviations from this shape. A good way to estimate the
importance of such various effects is to parameterize the observed BF
by means of a Gauss-Hermite orthonormal expansion, similarly to what
is described in \citet{VanDerMarel93} for the case of stellar
kinematics in elliptical galaxies. Following their method, if we call
$\alpha (v)$ the normal Gaussian function:
$$\alpha(v) = \frac{1}{\sqrt{2 \pi} \sigma_{v}} \exp \left(-
\frac{v^2}{2 \sigma_v^2} \right), \eqno(10)$$
where $\sigma_v$ is the line of sight velocity dispersion, we may
express the emission line BF as:
$$B(v) = B_0 \alpha(v - V_{sys}) \left[1 + \sum_{i = 3}^N h_i H_i(v -
  V_{sys})\right], \eqno(11)$$
where we call $B_0$ the BF normalization factor, $V_{sys}$ the systemic radial
velocity offset between the BF and the chosen reference frame, $H_i(v
- V_{sys})$ the $i^{\rm th}$ order Hermite polynomial, and $h_i$ the
corresponding coefficient. A wide description of the properties of the
Hermite polynomials is given in \citet{VanDerMarel93}. It is
demonstrated that odd order functions account for asymmetric
distortions of the Gaussian profile, while even order functions have a
symmetric effect. Truncating Eq.~(11) to $N = 4$, the Hermite
polynomials are expressed by: 
$$H_3(y) = \frac{1}{\sqrt{6}}(2\sqrt{2} y^3 - 3\sqrt{2} y) \eqno(12a)$$
$$H_4(y) = \frac{1}{\sqrt{24}}(4 y^4 - 12 y^2 + 3). \eqno(12b)$$
Therefore, it is possible to estimate the role of non-Gaussian
kinematical components, using the whole BF profile, simply by fitting
the observed shape with a truncated Gauss-Hermite series and measuring
the appropriate values of $h_3$ and $h_4$.

\begin{figure}[t]
\begin{center}
\includegraphics[width = 8.6cm]{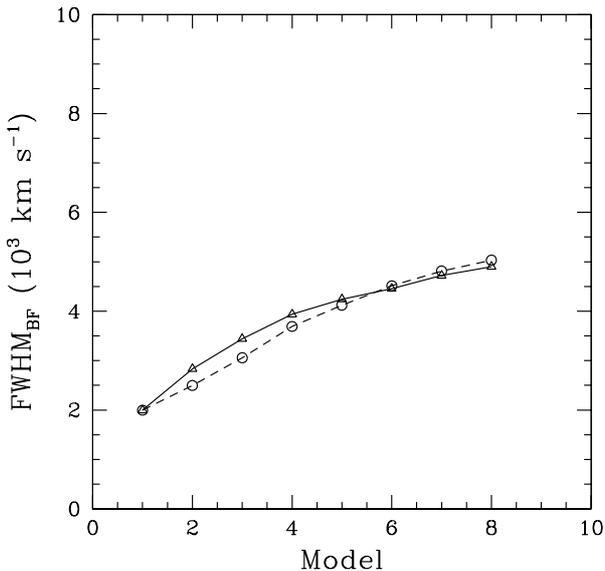}
\end{center}
\caption{Mass - inclination degeneracy. In this plot we illustrate the
  expected FWHM in the BF of emission lines originated in a disk
  structure. The empty triangles connected by the continuous line are the
  predicted values for reference disk models with $i = 10$° in the
  gravitational field of black holes with $M_{BH}$ = 1, 2, 4, 6, 8, 10, 12,
  and 14 $\cdot 10^7 {\rm M}_\odot$ (respectively models from 1 to 8);
  the circles with the dashed line show the predicted behavior for the
  reference disk model in the gravitational field of the black hole
  with $M_{BH}$ = $10^7 {\rm M}_\odot$, when seen under inclinations of $i =
  10$°, 15°, 20°, 25°, 30°, 35°, 40°, and 45°
  (again from model 1 to 8).}
\end{figure}
\subsection{The Balmer line broadening functions}
Applying our formalism to a set of BLR spectra, collected at the {\it
  Sloan Digital Sky Survey} (SDSS) spectroscopic database
\citep[see][for more discussion]{SDSSDR6, LaMura09}, our task is then
to recover their BF. To calculate the cross-correlation functions, we
build a template of Balmer emission lines, following the median line
intensity ratios found by \citet{LaMura07}. The template assumes that
the SDSS instrumental profile is a Gaussian function with FWHM $=
167\, {\rm km\, s^{-1}}$. At the spectral resolution of Sloan data,
the logarithmic sampling of the wavelength coordinate can be performed
with discrete bins corresponding to 69 ${\rm km\, s^{-1}}$ each. We
compute the template's ACF:
$$A(x) = T(x) \otimes T(x) \eqno(13)$$
and the cross-correlation functions of the BLR spectra with the
template, following the definition of Eq.~(7). Applying the least
squares formalism to the equation system~(9), it follows that the BF
of each spectrum must satisfy the relations:
$$\sum_{i = 0}^N B_i \left(\sum_{j = 0}^N A_j A_{i - k}\right) =
\sum_{i = k}^N A_{i - k} X_i. \eqno(14)$$

\begin{figure}[t]
\begin{center}
\includegraphics[width = 8.6cm]{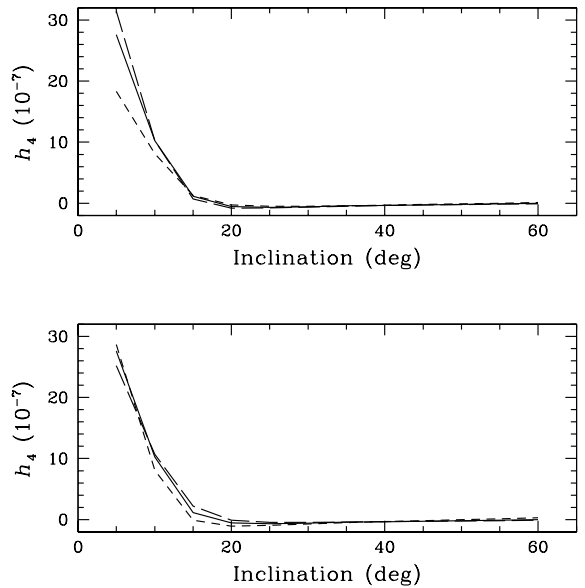}
\end{center}
\caption{The effect of inclination on the symmetric distortion of the BF. Here
  we show a comparison of the reference model predictions concerning the
  BF symmetric component (continuous lines) with four variants: in the
  upper panel we plot models with stronger (long dashed line) and
  weaker (short dashed line) disk emission with respect to the
  surrounding gas distribution; in the bottom panel we show the
  differences obtained by setting $\sigma_{Bell} = 0.07\, c$ (short
  dashed line) and $\sigma_{Bell} = 0.09\, c$ (long dashed line).}
\end{figure}
\section{BLR kinematical and structural models}
Since the BLR structure cannot be represented by a random motion pattern, the
shape of the broad emission lines may exhibit large deviations from the
Gaussian profile. If the BLR has a flattened component which is seen at low
inclination the geometrical structure can considerably affect the dynamical
interpretation of spectroscopic data. We illustrate this concept in Fig.~3,
where we plot the expected FWHM in the broadening function produced by disks
surrounding black holes of increasing mass and we compare it to the situation
of a black hole of fixed mass, but with the disk seen under different
inclinations.

Exploiting a combined BLR structural model, involving an accretion
disk and a surrounding distribution of gas \citep{Chenea89, Chen89,
  Popovic04}, we computed a range of emission line profiles, studying
the effect of source inclination. Our reference model assumes $R_{in} =
1834\, R_S$ for inner radius, $R_{BLR} = 18340\, R_S$ for outer radius,
$\sigma_{Disk} = 0.003\, c$ as the intrinsic velocity dispersion in the
disk, and $\alpha = -2.0$ for the radial emission power law, where
$R_S$ and $c$ represent the Schwarzschild radius and the speed of
light. Moreover, this model includes a surrounding gas distribution
having a velocity dispersion of $\sigma_{Bell} = 0.008\, c$. This model
carries out the best match to the observed line profiles with the
assumption of various disk inclinations.

In Fig.~4, we compare the reference model with some variants, obtained
with slightly different parameters. We note that all the models
predict a strong dependence of the symmetric non-Gaussian component (the
coefficient $h_4$ in the Gauss-Hermite expansion) on the disk
inclination, in the range of small values of $i$. The reason is quite
simple, because a nearly face-on disk enhances the low radial velocity
peak of the BF, while an edge-on disk is more likely to affect the high
velocity wings.

\begin{figure}[t]
\begin{center}
\includegraphics[width = 8.6cm]{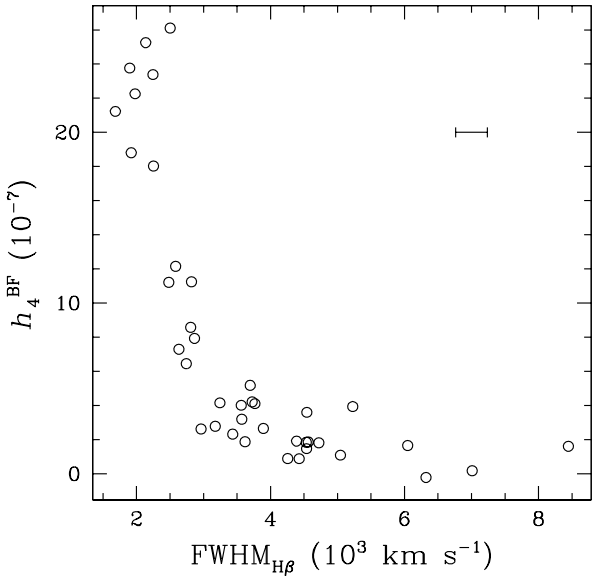}
\end{center}
\caption{BF symmetric distortion plotted as a function of FWHM$_{\rm
    H\beta}$. We observe a large evolution of the symmetric component as a
  function of broadening, suggesting the possibility of inclination effects in
  nearly face-on flattened structures. Such effects become weaker as
  the profile width increases.}
\end{figure}
As we show in Fig.~5, where we plot the measured values of $h_4$ as a
function of FWHM$_{\rm H\beta}$, there is a remarkable evolution of
the symmetric components of the line profiles, which decrease in importance
for increasing line profile width. Such an effect suggests that a
considerable variation of the geometrical factor $f$ might be present and it
should be taken into account in order to estimate the actual properties of
the SMBH. Using the model predictions, we can exploit the symmetric distortion
coefficient to estimate the inclination of the flattened BLR component
and to apply a correction to our dynamical interpretation of the
observed line profiles.

\begin{figure}[t]
\begin{center}
\includegraphics[width = 8.6cm]{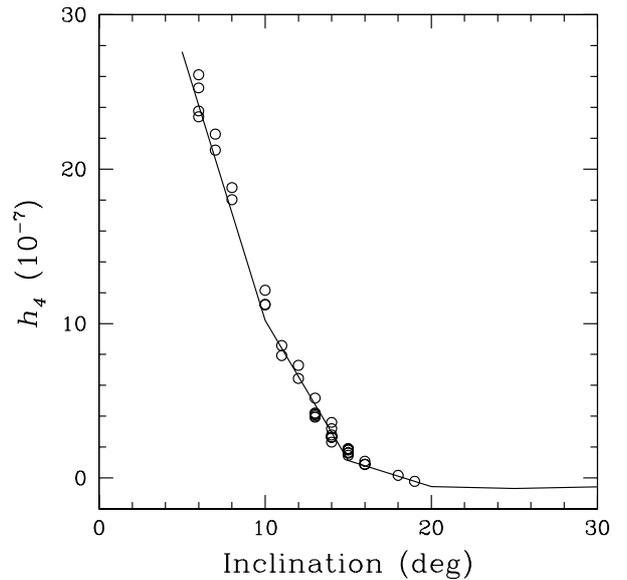}
\end{center}
\caption{BLR inclination inferred by comparison among the BF symmetric
  component $h_4$, predicted by the reference model for various inclinations,
  and the corresponding distribution observed in our data. According to the
  model predictions, the $h_4$ coefficient is very sensitive to flattening
  and inclination in nearly face-on structures, while it becomes a
  weaker indicator for larger inclinations.}
\end{figure}
\section{Results and discussion}
\subsection{The BLR geometry}
It can be shown that completely neglecting the role played by the BLR
geometrical factor may lead to incorrect black hole mass estimates, with
uncertainties that, in the worst cases, could span over two orders of
magnitude. Using the line profile analysis to estimate the BLR
structural properties, we apply a modified scheme to evaluate the
black hole mass in Eq.~(1), introducing an {\it equivalent velocity
  field}, such that $v_{eq} = f\, \Delta v$, defined as:
$$v_{eq} = \frac{1}{2}\left[\frac{\sqrt{3}}{2} {\rm FWHM}_{Bell}({\rm
    H}\beta) + \frac{{\rm FWHM}_{Disk}({\rm H}\beta)}{4\sin
    i}\right]. \eqno(15)$$
Assuming that the line profile broadening results from both planar and
non-planar motions \citep[][etc.]{Labita06, McLure02, Jarvis06},
$v_{eq}$ combines the velocity estimates obtained from the H$\beta$
emission line profile by fitting two Gaussian functions, which are
subsequently compared with the reference model, providing a
distinction among the disk and the surrounding gas contributions. The
corresponding geometrical factors are assumed to be given by the
isotropic interpretation of \citet{Netzer90} and the projection of a
rotating disk, confined in a smaller region with respect to the other
component. The inclination of the disk is estimated by comparison of
the BF symmetric distortion with that predicted by the reference model, as
shown in Fig.~6, while $R_{BLR}$ is inferred by measurements of the optical
continuum intensity, according to the empirical relationships found in some
recent reverberation mapping campaigns \citep{Kaspi00, Kaspi05, Bentz06}.

\begin{figure}[t]
\begin{center}
\includegraphics[width = 8.6cm]{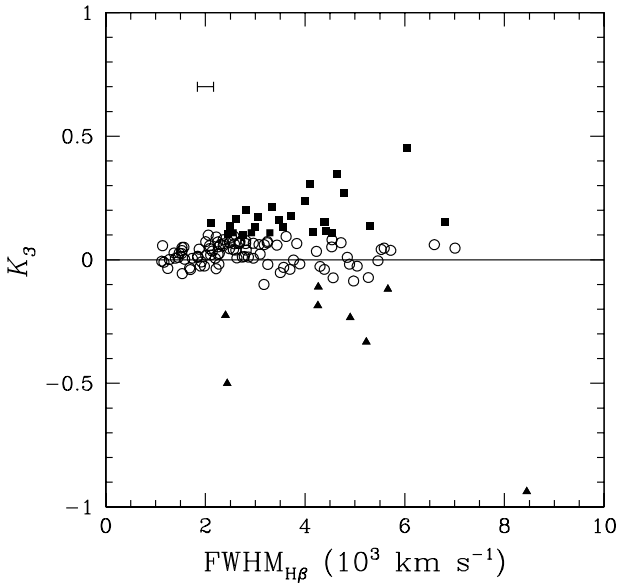}
\end{center}
\caption{H$\beta$ asymmetry parameter distribution vs. FWHM$_{\rm
    H\beta}$. Filled symbols represent objects where the asymmetric
  component exceeds 10\% of the Gaussian contribution. We plot as
  triangles objects that are affected by negative asymmetry, yielding
  red shifted peaks and blue shifted wings, while we use squares to
  represent positive asymmetry sources, with a blue shifted peak and
  red shifted wings. Large asymmetries characterize objects where
  fits with the reference model are more likely to be
  problematic. The bar in the upper left region of the diagram is a
  median estimate of the measurement errors.}
\end{figure}
A particularly important problem, involving the determination of AGN
physical properties from emission lines, resides in the line profile
asymmetries. Several factors, such as partial obscuration,
geometrical structure, or large scale non-virialized motions can
produce asymmetric line profiles. Moreover, relativistic effects
within the gravitational field of the SMBH give raise to asymmetries,
especially in the high velocity wings of the profile, which are
included in the calculations of the model by \citet{Chen89}. In order
to assess how much the asymmetric component affects our estimates of
the velocity field, we introduced an asymmetry parameter:
$$K_3 = h_3 H_3({\rm HWHM_{H\beta}}), \eqno(16)$$
expressing the relative contribution of the asymmetric component, with
respect to the Gaussian component, in the profile of H$\beta$ at its
half-maximum level. As we show in Fig.~7, the asymmetric component
gives a contribution to the FWHM which rarely exceeds the 10\%
level. The most extreme cases, where the asymmetric component becomes
larger than 20\%, occur only in the range of very broad line emitting
sources. Although this property is not particularly well represented in
our sample, it echoes the observation of larger asymmetries in objects where
${\rm FWHM_{H\beta}} > 4000\, {\rm km\, s^{-1}}$, which is among the features
identified by \citet{Sulentic00, Sulentic06} in their distinction between
Population A and B sources. Since there is evidence for different structural
configurations in objects with broader and narrower emission lines, at least
in the case of radio-loud sources, objects with the largest asymmetries are
more problematic in their comparison with the reference model.

\begin{figure}[t]
\begin{center}
\includegraphics[width = 8.6cm]{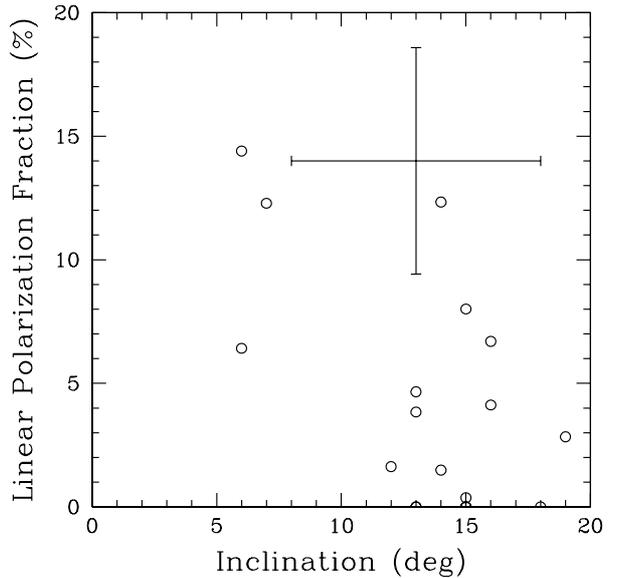}
\end{center}
\caption{Degree of linear polarization at the radio frequency of
  1.4~GHz as a function of the inferred BLR inclination. The cross in
  the upper left corner gives the median uncertainty
  estimate. Polarization data are from the NVSS catalogue.}
\end{figure}
\subsection{Implications}
Most of the results achieved in this work depend critically on the
choice of our reference model, which leads us to conclude that the
BLR has a flattened component, that is commonly seen at $i \leq
20$°. In the case of radio-loud sources, nearly face-on disk
structures are likely to produce a radio jet oriented along our line
of sight and the resulting signal should be considerably
polarized. Indeed, some objects of our sample have been detected in
the NRAO VLA Sky Survey (NVSS), which provides measurements of the
radio flux and polarization at the frequency of 1.4~GHz
\citep{Condon98}.\footnotemark \footnotetext{Polarization data are
  available at the web site
  http://www.cv.nrao.edu/nvss/NVSSlist.shtml} We compare the degree of
linear polarization with our inclination estimates in Fig.~8. We find
that a significant degree of linear polarization is detected in many
objects and it appears to be an averagely decreasing function of $i$.

\begin{figure}[t]
\begin{center}
\includegraphics[width = 8.6cm]{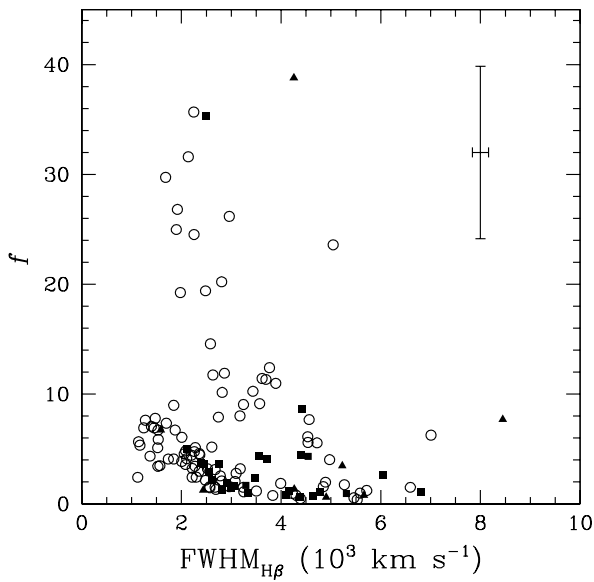}
\end{center}
\caption{The range of estimated geometrical factors for sources of
  different FWHM$_{\rm H\beta}$. Open circles represent objects with
  symmetric profiles, while filled squares and triangles are for
  sources with positive and negative asymmetries, respectively. The
  distinction among sources with different degree of asymmetry is
  required because objects with large asymmetries are harder to
  compare with the theoretical line profiles, predicted by the
  reference model.}
\end{figure}
A comparison of our mass determinations with the isotropic assumption
allows us to study the properties of the geometrical factor within our
sample. The situation depicted in Fig.~9 clearly indicates that
significant effects, up to a factor $\sim 40$, should be expected 
and that they are more commonly observed in the range of sources with
FWHM$_{\rm H\beta} \leq 3000 - 4000 {\rm km\, s^{-1}}$. We find that the
average value of the geometrical factor for black hole mass
determinations based on FWHM$_{\rm H\beta}$ is $f = 6.51$, not
very different from the result achieved by \citet{Onken04}, who gave
a geometrical factor $f = 5.5 \pm 1.9$ for black hole mass estimates based on
the emission line dispersions. We should note, however, that a previous
investigation by \citet{Collin06} applied composite kinematical models to the
BLR and compared several estimates of the central engine mass, suggesting
a distinction among the geometrical factors needed to correct
the velocity fields based on the emission line dispersion and FWHM.

Applying our modified scheme to the calculation of the central engine's
dynamical properties, we find that a correction for BLR inclination increases
our estimate of the SMBH mass. A direct consequence of this effect is that we
do not detect dramatically high accretion rates, with respect to the
corresponding Eddington limits. In particular, as it is illustrated in
Fig.~10, although we do observe a trend where objects with narrower emission
lines have typically higher accretion rates, this property does not appear to
hold in the form of a real anti-correlation, as it could look like, when the
BLR geometrical properties were not properly considered. Comparing our
estimates of accretion rate with the temperature parameters $A$, inferred by
means of the BP analysis and plotted in Fig.~11, we find that the occurrence
of high accretion rates is generally associated to lower values of $A$
(i. e. higher plasma temperatures), but we detect no particular relationship
with the effects that undermine the reliability of the BP method, suggesting
that more physical processes should be involved in this problem.

\begin{figure}[t]
\begin{center}
\includegraphics[width = 8.6cm]{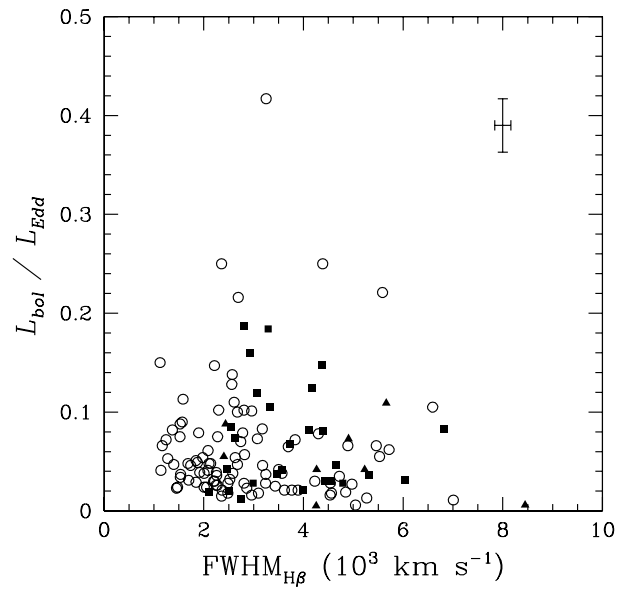}
\end{center}
\caption{SMBH accretion rates vs. FWHM$_{\rm H\beta}$, using the same
  symbology as in Fig.~7. The cross in the upper right corner of
  the diagram is the median uncertainty of measurements. We do not
  detect either very high accretion rates or systematic trends
  associated to the line profile widths. Our estimates do not appear
  to be dramatically affected by the presence of asymmetric profile
  components.}
\end{figure}
\begin{figure}[h!]
\begin{center}
\includegraphics[width = 8.6cm]{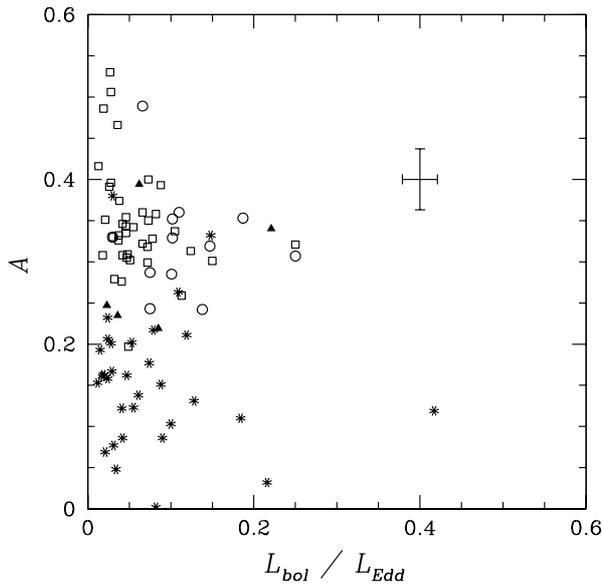}
\end{center}
\caption{The temperature parameter $A$ compared with the estimated accretion
  rate, in objects with different BP classification (symbols are the same as
  in Fig.~2). Lower values of $A$ are observed with increasing Eddington
  ratios.}
\end{figure}
\vspace{0.5cm}
\begin{center}
Acknowledgements
\end{center}
We thank the referee for useful discussion and suggestions concerning the
development of our research.

L. \v C. Popovi\'c was supported by the Ministry of Science of
R. Serbia through project 146002 ``Astrophysical Spectroscopy of
Extragalactic Objects.''

Funding for the SDSS and SDSS-II has been provided by the Alfred
P. Sloan Foundation, the Participating Institutions, the National
Science Foundation, the U.S. Department of Energy, the National
Aeronautics and Space Administration, the Japanese Monbukagakusho,
the Max Planck Society, and the Higher Education Funding Council
for England. The SDSS Web Site is http://www.sdss.org/. 

The SDSS is managed by the Astrophysical Research Consortium for
the Participating Institutions. The Participating Institutions are
the American Museum of Natural History, Astrophysical Institute
Potsdam, University of Basel, University of Cambridge, Case
Western Reserve University, University of Chicago, Drexel
University, Fermilab, the Institute for Advanced Study, the Japan
Participation Group, Johns Hopkins University, the Joint Institute
for Nuclear Astrophysics, the Kavli Institute for Particle
Astrophysics and Cosmology, the Korean Scientist Group, the
Chinese Academy of Sciences (LAMOST), Los Alamos National
Laboratory, the Max-Planck-Institute for Astronomy (MPIA), the
Max-Planck-Institute for Astrophysics (MPA), New Mexico State
University, Ohio State University, University of Pittsburgh,
University of Portsmouth, Princeton University, the United States
Naval Observatory, and the University of Washington.

\bibliographystyle{elsarticle-harv}

\end{document}